\documentclass[conference]{IEEEtran}
\IEEEoverridecommandlockouts
\usepackage{cite}
\usepackage{amsmath,amssymb,amsfonts}
\usepackage{algorithmic}
\usepackage{graphicx}
\usepackage{textcomp}
\usepackage{xcolor}
\usepackage{tikz}
\usepackage{pgfplots}
\usepackage{soul}
\usepackage{adjustbox}
\usepackage{diagbox}
\usepackage{enumitem}
\usepackage{microtype}
\usepackage{graphicx}
\usepackage{amsmath}
\usepackage{amssymb}
\usepackage{pifont}
\usepackage{algorithm2e}
\usepackage{color}
\usepackage{colortbl}
\usepackage{multirow}
\usepackage{xcolor}
\usepackage{booktabs}
\usepackage{floatrow}
\usepackage{tabularx}
\usepackage{xurl}
\usepackage{subcaption}
\usepackage{dblfloatfix}
\definecolor{LightCyan}{rgb}{0.88,1,1}
\definecolor{lightgoldenrodyellow}{rgb}{0.98, 0.98, 0.82}
\definecolor{lightkhaki}{rgb}{0.94, 0.9, 0.55}
\definecolor{lightyellow}{rgb}{1.0, 1.0, 0.88}

\def\BibTeX{{\rm B\kern-.05em{\sc i\kern-.025em b}\kern-.08em
    T\kern-.1667em\lower.7ex\hbox{E}\kern-.125emX}}

\begin{document}
	
	\title{COPS: A Compact On-device Pipeline for real-time Smishing detection}

		\author{
			\IEEEauthorblockN{Harichandana B S S}
		\IEEEauthorblockA{\textit{Samsung R\&D Institute} \\
			Bangalore, India \\
			hari.ss@samsung.com}
			\and
		
		\IEEEauthorblockN{Sumit Kumar}
		\IEEEauthorblockA{\textit{Samsung R\&D Institute} \\
			Bangalore, India \\
			sumit.kr@samsung.com}

		\and
		\IEEEauthorblockN{Manjunath Bhimappa Ujjinakoppa}
		\IEEEauthorblockA{\textit{Samsung R\&D Institute} \\
			Bangalore, India \\
			manjunath.bu@samsung.com}
	
		\and
		\IEEEauthorblockN{Barath Raj Kandur Raja}
		\IEEEauthorblockA{\textit{Samsung R\&D Institute} \\
			Bangalore, India \\
			barath.kr@samsung.com}
		
	}

		
		\maketitle
		
		\begin{abstract}
Smartphones have become indispensable in our daily lives and can do almost everything, from communication to online shopping. However, with the increased usage, cybercrime aimed at mobile devices is rocketing. Smishing attacks, in particular, have observed a significant upsurge in recent years. This problem is further exacerbated by the perpetrator creating new deceptive websites daily, with an average life cycle of under 15 hours. This renders the standard practice of keeping a database of malicious URLs ineffective. To this end, we propose a novel on-device pipeline: COPS that intelligently identifies features of fraudulent messages and URLs to alert the user in real-time. COPS is a lightweight pipeline with a detection module based on the Disentangled Variational Autoencoder of size 3.46MB for smishing and URL phishing detection, and we benchmark it on open datasets. We achieve an accuracy of 98.15\% and 99.5\%, respectively, for both tasks, with a false negative and false positive rate of a mere 0.037 and 0.015, outperforming previous works with the added advantage of ensuring real-time alerts on resource-constrained devices.
		\end{abstract}
		
		
		\section{Introduction}

		The number of smartphones globally has surpassed six billion, according to research firm Statista \cite{statistica}. It is expected to increase by several hundred million more over the next several years. Consequently, phishing attacks on smartphones are on the rise as they serve as a lucrative opportunity for cyber attackers to steal user information. Mobile users falling for phishing attacks have increased by 160\% Year-over-Year \cite{jamf}. These attacks cause the consumer to lose millions of dollars yearly in the US alone \cite{tableu}.
		


		Smishing, as the name sounds, is the combination of two words SMS (Short Message Service) and Phishing. With the rise of Over-The-Top (OTT) platforms like Whatsapp, Telegram, etc., phishing messages have expanded their reach on these platforms. According to the statistics \cite{kaspersky}, Kaspersky Internet Security for Android detected that the most significant share of detected malicious links between December 2020 and May 2021 was sent via WhatsApp (89.6\%), followed by Telegram (5.6\%). Smishing attacks frequently request that the victim open a link, call a number, or send an email address the attacker has provided through an SMS message.

		\begin{figure}[t]
		\centering
		\includegraphics[width=\linewidth, interpolate=false]{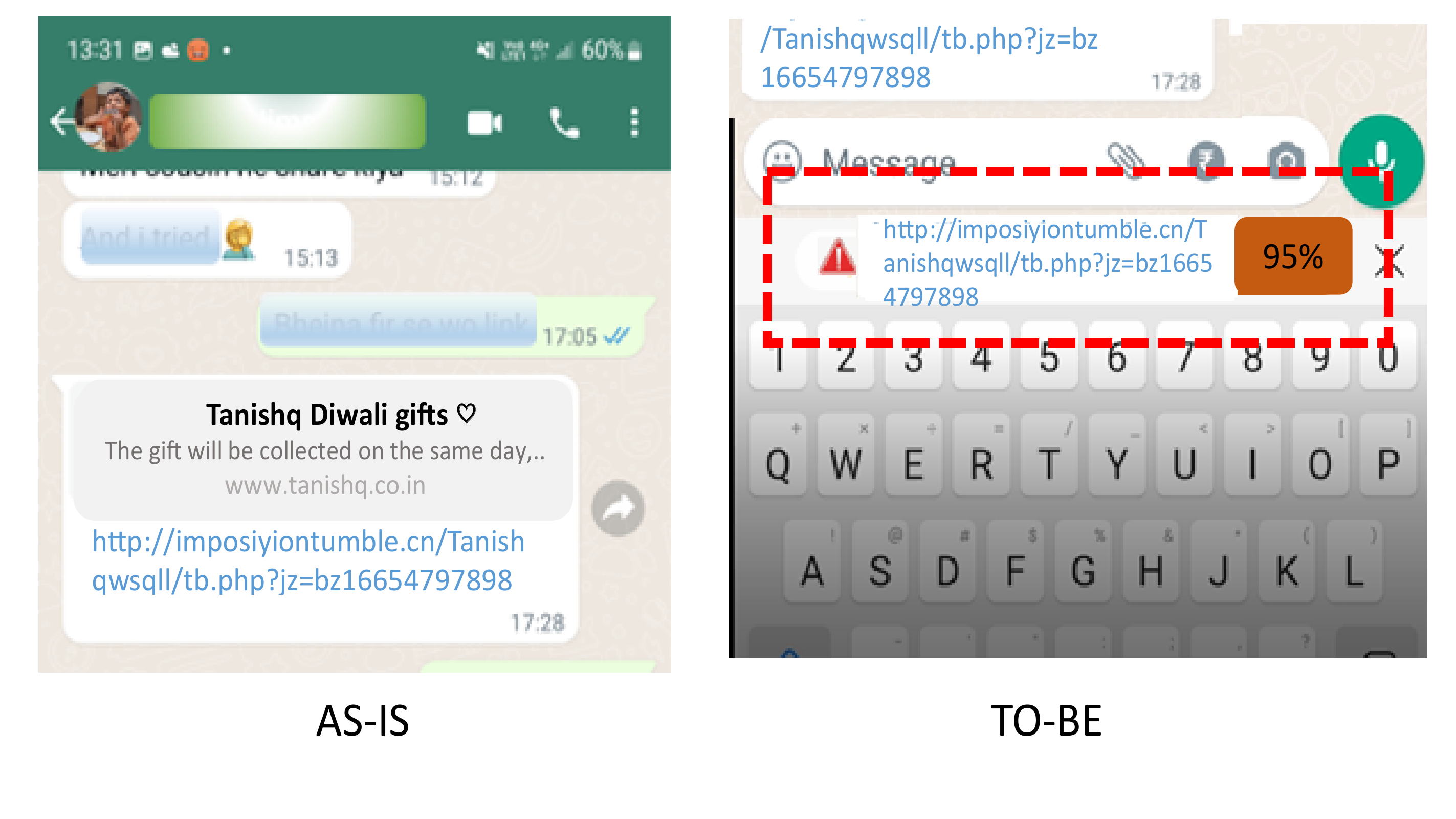}
		\caption{Smishing Detection: To the left is the existing system, and to the right is a system with the COPS pipeline}
		\label{fig:smishdetect}
		\vskip -0.20in
		\end{figure}

		Research conducted by \cite{lifecycle} observed that the bulk of phishing pages was only active for less than 24 hours. This yields the block-list-based approaches to be ineffective. Some works tackle this issue using more complex architectures. However, deploying such models in real-time and on-device is challenging. Another critical issue with existing Smishing detecting systems is their integration with Smartphone OS. Most of the time, separate applications are developed to detect such messages, which are not effective. It is inconvenient for the user to select and verify in another application whenever any message is received. The natural tendency is to click the links in the messages as soon as it arrives \cite{jamf}. Adding to this, new smishing techniques utilize short URLs, which may redirect to a malicious link upon clicking.
		
		Autoencoders have proven to perform well for tasks like Anomaly detection \cite{anomaly}. But, one drawback of these is that the latent space needs to be regularized. Variational Autoencoders followed by $\beta$-VAE were introduced to solve this. However, works on real-time smishing detection using $\beta$-VAE are scarce. We thus introduce a Compact On-device Pipeline for real-time Smishing detection (COPS) based on the $\beta$-VAE. 
		
		Figure \ref{fig:smishdetect} overviews the existing system and a system with COPS implemented. Many smishing messages come with a known brand name or associated icons. However, instead of the original website link, a phishing link is provided. Aware of the brand icon, the user clicks the link and gets phished. Instead, if a COPS is deployed, the user can be alerted in real-time about the potentially dangerous nature of the message, and he/she can make an informed decision.

		In this paper, our main contributions are as follows:
		\begin{itemize}
			\item  We propose, for the first time, a Compact On-device Pipeline for real-time Smishing detection (COPS) optimized to deploy on resource-constrained devices and inform the user regarding Smishing without leaving the app on which the user is communicating. We achieve this with a total memory footprint of a mere 3.46MB.
			
			\item We benchmark COPS on smishing and URL Phishing detection against open datasets and report the results. We further propose a pipeline for synthetic data generation.
			
			\item We also show that our proposed model outperforms prior works in phishing and smishing detection tasks.
		\end{itemize}

		
		
		\section{Related Works}
		Studies have shown different approaches to detect smishing attacks. Some popular methods include Block-listing and Heuristic approach,  Rule-based, feature-based with Machine Learning (ML), and deep learning (DL) techniques.
		
		Block-listing is a common and classical way to detect malicious URLs where we continuously add spam, malware, or smishing URLs to a block list. One such work which describes a block-list-based solution is  \cite{hong2020phishing}. \cite{gupta2018defending} described a white-list-based strategy in which many characteristics of genuine websites were noted. However, one major drawback of this is that It is tedious to maintain and needs continuous upgrading. Thus, there is a high chance of missing out on unknown URLs. 
		
		A rule-based approach was presented in \cite{moghimi2016new}. Other rule-based research works include  \cite{jain2018rule} \cite{foozy2014practical}. Following are some advanced ways of detecting smishing URLs.
		
		Heuristic approaches are an extension of Block-list methods. A heuristic-based approach was suggested \cite{mishra2021dsmishsms}, in which 20 heuristic traits were chosen. The outcomes demonstrate the superiority of the heuristics over the blacklist-based method.

		To further improve the accuracy from rule-based, \cite{li2020improving} suggests using linear and non-linear transformation methods by extracting valuable features from each URL. A methodology to detect phishing websites in real scenarios is showcased in the study \cite{sanchez2022phishing}. Some prior works like \cite{kaya2016novel} and \cite{jain2019feature} use feature-based techniques for smishing.

		Due to the advent of a comprehensive range of ML techniques, studies on ML-based smishing detection are vast. A study \cite{boukari2021machine} shows the usage of machine learning techniques to detect phishing and Smishing frauds. On similar lines, \cite{jain2020novel} utilizes Ham and Spam messages to detect smishing attacks.	Another technique is explored by \cite{pattewar2019malicious}, where the method tries to analyze the URL details and its websites.

		An embedding-based approach, \cite{sen2021malicious} used Convolution Neural network (CNN) and Long Short Term Memory (LSTM). They show that hybrid models provide better accuracy than individual models, and Character-embedded models perform better than word-embedding models.
		
		The Autoencoder (AE) was first formally defined by Baldi \cite{baldi2012autoencoders} and aimed to reconstruct given input. Liou et al. \cite{liou2014autoencoder} proposed an Autoencoder for text. Some works on AE-based anomaly and phishing detection are \cite{lin2020anomaly} \cite{zhang2021multiphish} \cite{metlapalli2020classification} \cite{sharaff2022deep}. Many studies have focused on further improving AE. Variational Autoencoders (VAE) \cite{kingma2013auto} and $\beta$-VAE \cite{higgins2016beta} are among these. 
		
		The methodologies mentioned above can realize their potential only if they can be deployed in real-time in end-user devices. Existing deployment techniques require manual effort, which is a huge hassle. To overcome this, we plan to use a novel pipeline with a Screen Understanding module inspired by \cite{murthy2022dsu} (explained in Section \ref{sec:sysdesign}), independent of the underlying application, followed by a Smishing detection network. Furthermore, very few/no studies exist on real-time Smishing detection using $\ beta$-VAE-based architectures. 
		
		\section{System Design}
		\label{sec:sysdesign}

\begin{figure}[h]
		 \centering
		  \includegraphics[trim={0cm 9cm 0cm 9.3cm},clip, width=\textwidth, interpolate=false]{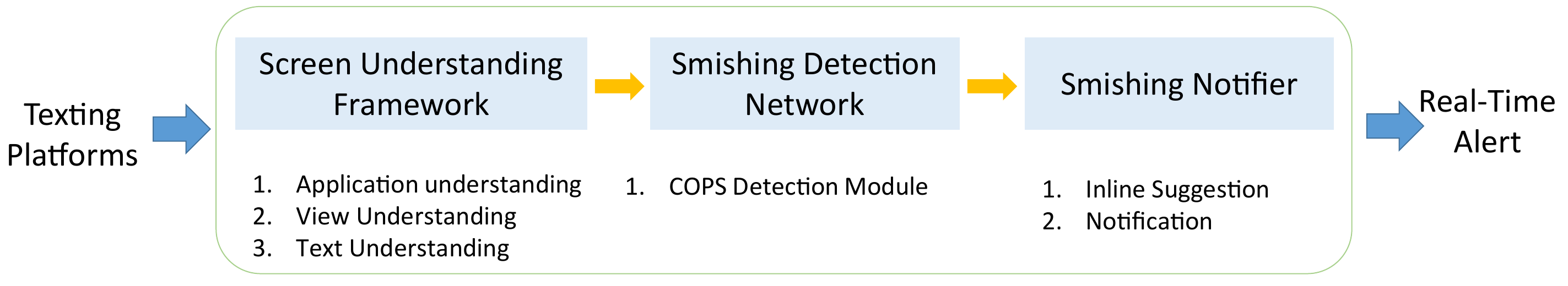}
		  	\caption{COPS System Design: for a real-time \textit{Smishing} detection}
			\label{fig:system_design}
	
		\end{figure}

		Most existing Smishing detection solutions require users to manually copy the received SMS to a different application and check its authenticity. This method is not convenient as many manual steps are involved, and sometimes messages are pretty convincing to trap using in clicking the smishing link. Another approach is to change the default app and use the application with built-in phishing detection. However, changing the favorite message application may not seem valid just for detecting smishing. Also, in the current scenario, users are on multiple platforms, which renders the above method ineffective. 
		To this end, we propose an end-to-end smishing detection pipeline independent of underlying messaging platforms. The pipeline shown in Figure \ref{fig:system_design} consists of three major modules:
		\begin{itemize}
			\item Screen Understanding framework: Screen understanding framework has three major functionalities: 
			\begin{itemize}
				\item Understanding which application is currently displayed, based on the understanding, process just the screen related to the communication/messaging category.
				\item  Understand the views or user interface (UI) elements associated with the displayed screen.
				\item Extract the texts related to views
			\end{itemize}

			\item Smishing Detection network: Verify the textual content provided by the Screen understanding framework. The model architecture is explained in detail in Section \ref{sec:modelDes}.
			\item Smishing Notifier: Once Smishing is detected with high probability, the smishing notifier can notify the user by providing notifications or inline suggestions.
		\end{itemize}

		\section{Dataset Specifications}
		 Smishing and Url Phishing detection problems depend highly on data. One major challenge in dataset identification/collection is that due to the short-lived nature, and creation of new malicious URLs, there is an absence of benchmark datasets. However, due to extensive research in this field, some publicly available datasets are regularly updated. Furthermore, we also explore synthetic data generation for smishing detection to help further improve model performance.	This section discusses the utilized datasets and the pre-processing techniques involved.
		
		\label{sec:DataCollection}
		\subsection{Task Specific Dataset Details}
		
		\subsubsection{Smishing}
		\label{sec:DataSmishing}
		We use the dataset described in \cite{mishra2022implementation} for smishing detection \cite{smishingdata}. The detailed distribution is shown in Table~\ref{tab:smishdata}. We can observe that the data is highly imbalanced. We solve this issue by using class weights while training. Further, we also use another open dataset \cite{kaggle} for testing our model.
		
		\begin{table}[h]
			\caption{Smishing dataset \cite{smishingdata} details}
			\label{tab:smishdata}
			\centering
				\resizebox{0.8\linewidth}{!}{
			\begin{tabular}{cccc}
				\toprule
				\rowcolor{lightyellow}
				Category & \# Total Samples & Train Set & Test Set\\
				\midrule
				Ham & 4844 & 4328 &  516\\
				Spam & 489 & 454 &  35\\
				Smishing & 638 & 592 & 46  \\
				\bottomrule
			\end{tabular}}
				\vskip -0.09in
		\end{table} 
		\subsubsection{Generated Data}
		As the smishing dataset is highly imbalanced and small, we generate synthetic data using the $\beta$-VAE model on the train split. The details of the model used for data generation are explained in Section \ref{sec:modelDes}. Table \ref{tab:datagen} shows some synthetically generated data samples. We increase the samples of Spam and Smishing classes to two times the initial amount. 
		
					\begin{table}[h]
		\caption{Generated data samples}
		\label{tab:datagen}
		\resizebox{0.9\textwidth}{!}{
		\begin{tabular}{p{0.47\textwidth} p{0.47\textwidth}}
			\toprule
			\rowcolor{lightyellow}
			Reference Sentence  & Generated Sentence \\
			\midrule
			Please CALL 08712402779 immediately as there is an urgent message waiting for you  & please call  *************** immediately as there is an urgent message waiting for you \\
			\midrule
		FREE for 1st week! No1 Nokia tone 4 ur mob every week just txt NOKIA to 8007 Get txting and tell ur mates www.getzed.co.uk  & free for 1 st week ring tone nokia txt [oov] \\
			\bottomrule
		\end{tabular}}
			\vskip -0.09in
	\end{table}
		
		\begin{table}[h]
			\caption{URL Phishing dataset details}
			\label{tab:urlhdata}
				\resizebox{\linewidth}{!}{
			\begin{tabular}{cccc}
				\toprule
				\rowcolor{lightyellow}
				Category & \textbf{$Dataset_{1}$ \cite{malurl}} & \textbf{$Dataset_{2}$ \cite{faizanurl}} & \textbf{$Dataset_{3}$ \cite{phishstorm}} \\
				\midrule
				Not-Phishing & 428,103 & 344,828  &  48,009 \\
				Phishing & 94,111 & 75,643 & 47,904  \\
				\bottomrule
			\end{tabular}}
				\vskip -0.09in
	
		\end{table}
		\subsubsection{URL Phishing}
		We use datasets from three different sources for the task of URL Phishing: \cite{malurl} \cite{faizanurl} \cite{phishstorm}.
		The details of each dataset are shown in Table \ref{tab:urlhdata}.
		
		We test the model performance by two methods explained in detail in section \ref{sec:exp}. The results give insights into how the model performs on samples not part of a seen distribution (dataset).  
		
		
		\label{sec:DataPhishing}
\begin{figure*}[tb]
	\centering
	\includegraphics[trim={0 0 0 0},clip, width=0.65\linewidth, interpolate=false]{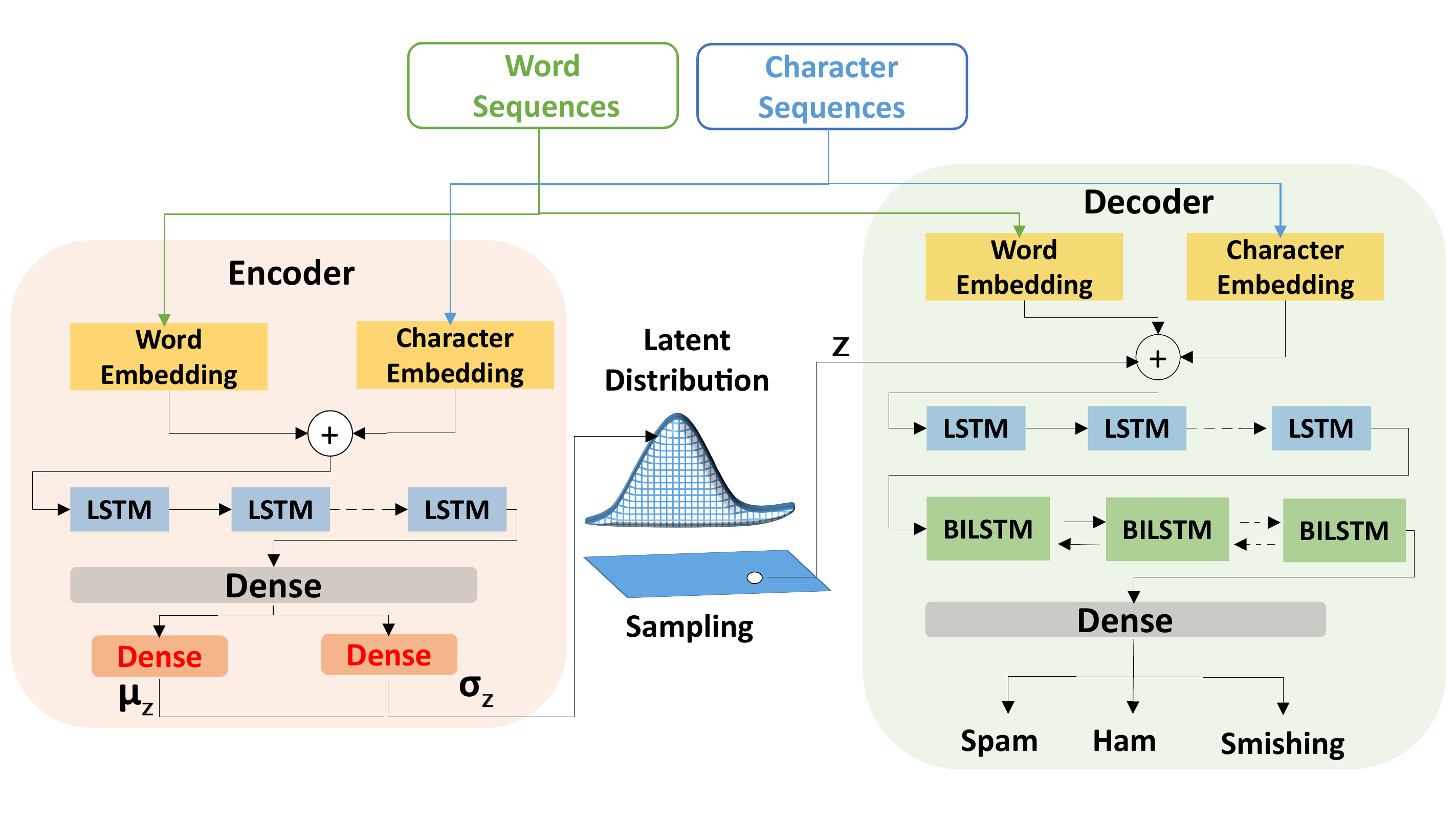}
	\caption{The proposed architecture for COPS detection module}
	\label{fig:ADVAnS}
	\vskip -0.28in
\end{figure*}
		
		\subsection{Data Pre-processing}
		Data pre-processing is the preliminary step required to achieve good performance. Mainly four steps are involved: Cleaning, Short URL Expansion,  Tokenization, Sequence Creation, and  Padding. Since smishing messages also contain URLs, and short URLs are most commonly used, it is essential to expand these without actually opening the link. To do this, we utilize java.net.HttpURLConnection. During the tokenization process, we create a word and character vocabulary of the required sizes (explained in Section \ref{sec:impdetails}).
		

		\section{Model Description}
		\label{sec:modelDes}

		Smishing attacks aim to extract user-private information like credit card access, bank details, etc., by either providing malicious URLs or convincing users to take action (call back, message, etc.) through the textual content of the message. Thus, detecting smishing attacks is challenging \cite{mishra2021dsmishsms}, considering that features of both textual and embedded URLs are equally important. To this end, we formulate the COPS detection module, an architecture that utilizes the features extracted by the  $\beta$ Variational Autoencoder \cite{higgins2016beta}. We also propose a module for data generation (Section \ref{sec:DataGen}) to improve smishing detection further.

		\subsection{Encoder}
		\label{sec:modelDesEnc}
		The encoder module of the $\beta$-VAE is responsible for compressing the input data into a latent space ($Z$) of dimension significantly smaller than the input. Let $I$ represent the input data, then the output of the encoder can be a latent space $Z$ by learning and optimizing certain parameters $\theta$. The encoder can be represented as $E_{\theta}(Z|I)$. 
		
		For the task of Smishing and Phishing prediction, the input is given in the form of `Word sequences' ($\vec{W}$) and `Character sequences' ($\vec{C}$) obtained after pre-processing data phase as shown in Figure \ref{fig:ADVAnS}. These inputs are then passed to the embedding layers, after which they are concatenated:
		\begin{equation}
			\small
			\begin{aligned}
				WC & = [\vec{W_{e_1}}, \vec{W_{e_2}}, \cdots, \vec{W_{e_m}}] \quad {\Big\Vert} \quad [\vec{C_{e_1}}, \vec{C_{e_2}}, \cdots, \vec{C_{e_n}}]\\
			\end{aligned}
		\end{equation}
		Where $\vec{W_{e_{i}}}$ is the word embedding vector of the $i^{th}$ word token in $\vec{W}$ and $\vec{C_{e_{j}}}$ is the character embedding vector of the $j^{th}$ character token in $\vec{C}$ and $WC$ is the result after concatenation (${\Vert}$). The next layer is the LSTM, which outputs the extracted feature representations from the hidden states.	This is followed by two Dense layers with a Latent dimension for extracting both $\mu_{Z}$ and $\sigma_{Z}$ to fit a latent distribution (Z).

		\subsection{Decoder}
		The decoder is the next half of the model, which utilizes the feature representations from the encoder and learns to decode them to arrive at the final output. We aim to develop decoder architectures for two main tasks in this paper data generation and classification. We design the decoder to sample from the latent distribution from the encoder and learn to optimize performance on the two mentioned tasks. The decoder design for each task is explained in the following sections. 
		\subsubsection{Sampling}
		The first step for the decoder, regardless of the task being data generation or classification, is sampling. The decoder is given a sample from latent distribution to reconstruct the input. The sampling procedure is straightforward. A random sample is drawn from the distribution G($\mu_{z}$, $\sigma_{z}$). However, for the convenience of backpropagation during training, sampling is taken from $\mu_{z} + exp^{\sigma_{z}} + \epsilon$ where $\epsilon \in G(0,1)$ which is equivalent to sampling from the G($\mu_{z}$, $\sigma_{z}$) distribution. 
		
		\subsubsection{Decoder for Data Generation}
		\label{sec:modelDecGen}
		We design the decoder to reconstruct the original Input `$I$' given a sample from the latent distribution `Z'. Let us define `$\gamma$' as the decoder weights and bias for the generation task. Learning and optimizing these parameters lead to the decoding process. The data generation decoder can be represented by `$D_{\gamma}(I|Z)$'. 
		
		Once the decoder is trained, we generate synthetic data using references from the original data. Let $S_{1}$ be a reference sentence in the data `D', then we first find another sentence $S_{2}$ in the dataset with the closest cosine similarity with $S_{1}$ and find the shortest homology for these inputs. Table \ref{tab:datagen} shows some samples for the generated sentences.

		\label{sec:DataGen}

		\subsubsection{Decoder for Smishing and URL Phishing}
		\label{sec:modelDecSmish}
		The decoder tries to extract essential features of the given Input `$I$' given a sample from the latent distribution `Z' obtained from the encoder. The decoder can be represented by `$D_{\alpha}(I|Z)$'. We design the decoder to predict the probability distribution of each label (`HAM', `SPAM', and `SMISHING'; for URL Phishing, we have tags: `PHISHING' and `NOT PHISHING') and thus act as a classification architecture.
		
		For the Smishing and Phishing detection task, we use word sequences, character sequences, and sampled parameters from the latent distribution as the inputs. The word and character sequences are first embedded using embedding layers and then concatenated with the repeated vectorized for of the parameter `Z'. These are then given to the LSTM layer followed by a BILSTM layer, as shown in Figure \ref{fig:ADVAnS}.
		\subsection{Loss}
		\label{sec:Loss}
		The loss function of the $\beta$-VAE can be decomposed into two parts: reconstruction loss and the KL divergence. The $\beta$-VAE loss function is the weighted sum of the two losses.
		
		\subsubsection{Data Generation Loss}
		As explained in Section \ref{sec:modelDesEnc} and Section \ref{sec:modelDecGen}, the encoder for the task of data generation can be represented by $E_{\theta}(Z|I)$ and the decoder can be represented as $D_{\gamma}(I|Z)$. The reconstruction loss is then:
		\begin{equation}
			\small
			\begin{aligned}
				Loss_{recon} & = - \mathbb{E}_{E_{\theta}} [log D_{\gamma}(I_{i}|Z)]  \\
			\end{aligned}
		\end{equation}
		It is the negative log-likelihood of the data points and nothing but the Mean Squared Error (MSE) of the original and reconstructed input. 
		The KL-divergence can be expressed as:
		\begin{equation}
			\small
			\begin{aligned}
				Loss_{KL} & = \sum_{i=1}^{N}exp^{\sigma_{Zi}} +  \mu_{Zi}^{2}  -  \sigma_{Zi}  - 1 \\
			\end{aligned}
		\end{equation}
		where P(Z) is the true distribution, $\sigma_{Z}$ and $\mu_{Z}$ are the encoder-fitted latent distribution representatives. Thus, this represents the error in fitting the latent distribution of the data points and measures the total loss of information.
		
		The total loss for data generation is given by:
		\begin{equation}
			\small
			\begin{aligned}
				VAE\_Loss_{generation} & = Loss_{recon} + \beta*Loss_{KL}  \\
			\end{aligned}
		\end{equation}
		The value of $\beta$ is greater than 1 for the model to represent a Disentangled Variational Autoencoder. It is set to be 74 after hyper-parameter tuning explained in Section \ref{sec:impdetailsHyp}.
		
		\subsubsection{Smishing and URL Phishing Loss}
		
		The loss for this task is similar to the data generation loss with the decoder taken to be $D_{\alpha}$, and the reconstruction Loss is taken on the predicted labels with the original labels to fit the classification model. Along with the VAE Loss, we also use a categorical-cross-entropy loss with class weights to address the class imbalance issue in the dataset.

		\section{Implementation Details}
		\label{sec:impdetails}
		We use Keras \cite{gulli2017deep} with Tensorflow \cite{abadi2016tensorflow} as a backend for model scripts. The models are trained using The NVIDIA GeForce GPU (358 GTX 1080 Ti with 12GB RAM or memory).
		
		\subsection{Hyper-parameter Tuning}
		\label{sec:impdetailsHyp}
		Hyper-parameter tuning can bring out the most from a model architecture. We use the Grid-Search technique for hyper-parameter tuning of COPS. We define tunable parameters to be: learning rate, batch size, latent dimension, LSTM, and BILSTM dimensions, $\beta$ in VAE Loss, and the optimizer for the model. Some essential results from hyper-parameter tuning are discussed in Section \ref{sec:results}. The details of the hyper-parameters chosen are as follows.
		
		\subsubsection{Data Generation}
		We set the Character vocabulary size as 80 and the Word vocabulary as 8,427. The embedding dimension for word and character sequences is set to 50. The encoder LSTM has a dimension of 64, the latent dimension is set to 32, and the first dense layer has a size of 96. After hyper-parameter tuning, the beta value is 74 for the VAE Loss calculation. We train the model using the Adam optimizer (lr = 0.001).

		\subsubsection{Smishing and URL Phishing prediction}
		The parameters for Smishing are similar to that of data generation with minor changes. We use a character vocabulary of 31 and a word vocabulary of 30,000 for the URL phishing prediction task. The latent dimension is set to 2. The decoder LSTM has a dimension of 100, a dropout of 0.5, and the BILSTM has a dimension of 50.

		\section{On-device Deployment for real-time Threat assessment}
		\label{sec:on-device}
		\begin{figure}[tb]
			\centering
		\includegraphics[trim={0 0 0 0},clip, width=\textwidth, interpolate=false]{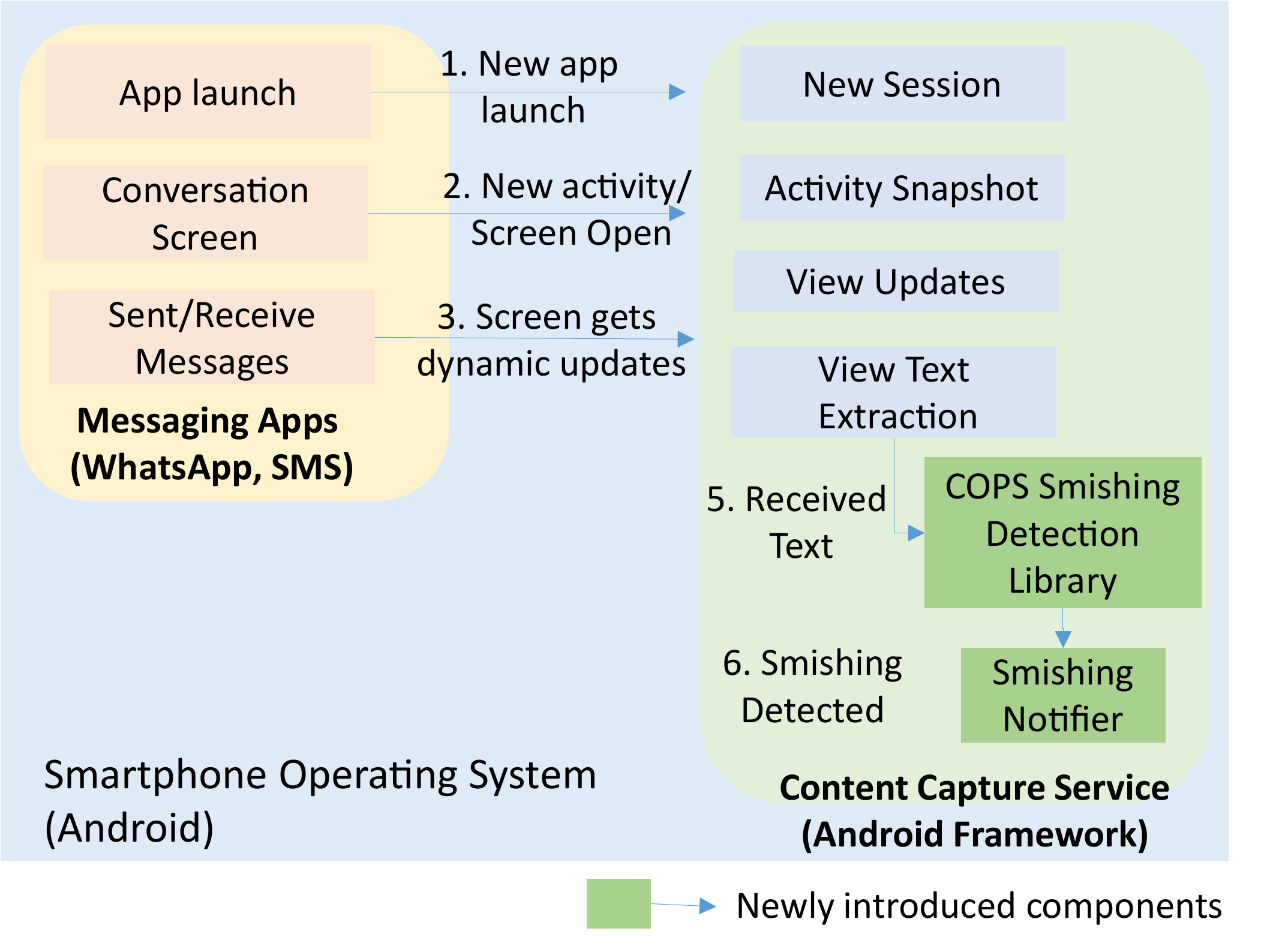}
			\caption{On-device real-time Smishing detection flow for Android}
			\label{fig:deployment_android_flow}
			\vskip -0.15in
		\end{figure}
		We have chosen Android as a mobile platform to demonstrate the working of COPS owing to its large-scale smartphone market share. The pipeline for smishing detection specific to Android is shown in Figure \ref{fig:deployment_android_flow} and explained below.
		
		\subsection{Screen Understanding Framework}
		The Screen Understanding Framework automatically captures relevant content in real-time from an application screen. The framework is further optimized for the efficient working of the proposed architecture of the COPS detection module. The details of the framework are explained in the following sections.
		
		\subsubsection{Real-time Content Capture}
		Android Screen Understanding framework has a content capture service (CCS) component. Content capture makes a real-time, continuous recording of application activity, display, and events possible. Whenever the user launches an activity inside an application, the Android framework provides a new snapshot for that activity to CCS. A View occupies a rectangular area on the screen and is responsible for drawing and event handling \cite{Android}. The complete view hierarchy of the screen consisting of all the views is maintained in a view node which will have information about parents' and children's views. We focus on the views containing text messages from the hierarchy and extract their content using the Text capture API provided by the Android framework. We temporarily cache the extracted text without sharing it with any other app or server for these views. Once the textual content is verified for phishing, the same is discarded.
		
		\subsubsection{Optimization}
		Invoking CCS for all screens and further verification can lead to unnecessary power drain in the devices. To optimize the working of COPS and invoke it only for the relevant category of applications, we maintain a safelist of well-known communication-related apps like Whatsapp, SMS, etc., region-wise. This is identified through a set of metadata specified by the application developer while uploading the application to the app store. Here region is essential, as different messenger apps have prominence in different countries, e.g., KakaoTalk is famous in Korea, while WeChat is comparatively widespread in China \cite{korea} \cite{china}. Also, the system analyzes all the screens of the application, and screen categorization is done to optimize the working of COPS further.

		\subsection{Smishing Detection Network}
		COPS detection deep neural model, with a size of 3.46 MB, is deployed on the device using the lightweight version of Tensorflow available for mobile called TensorFlow lite. The pre-processing and model inference logic is encapsulated in a library for easier access in the Android Archive (AAR) file form. The AAR file receives a text from the screen understanding module and tries to classify it.

		\subsection{Smishing Notifier}
		
		Once the system detects a Smishing attack on the user, it is equally important to notify the user before the URL gets selected or interacted with. We have explored two user interface (UI) options specific to Android below to gain user attention.
		
		\subsubsection{Inline Suggestion} The most interacted UI element in a conversation screen is the Input method editor(IME) or soft keyboard. When the user taps on edited text, Content Capture Service gets a fill callback, a connection to IME. This established connection can provide extra information to IME in the form of Inline suggestions that appear on top of the keyboard. Owing to its easier discoverability, the user can get immediately notified, as shown in Figure \ref{fig:system_design}.
		
		\subsubsection{Notification} Another way users can be notified is through Notifications. These are asynchronous events and can be updated in the background. Once the smishing detection system identifies a Smishing message, we can notify the user about the detected URL and application details.

		\section{Results and Analysis}
		\label{sec:results}
		The following subsections explain the experimental setup process and performance analysis of COPS.
		
		\subsection{Experimental Setups}
		\label{sec:exp}
		To evaluate our model, we test the performance in different experimental setups for phishing and smishing detection tasks. The experimental setups are described as follows:
		
		\textbf{\textit{Setup 1: URL Phishing} (S1):} In this setup, we utilize all three phishing datasets described in section \ref{sec:DataPhishing} for training and testing the model. $COPS_{mixed\_test}$ represents the model trained using this setup. We evaluate and compare $COPS_{mixed\_test}$ with existing works using a similar setup.

		\textbf{\textit{Setup 2: URL Phishing} (S2):}  This setup utilizes only two datasets i.e datasets \cite{malurl}, and \cite{faizanurl}. The third dataset \cite{phishstorm} is completely unseen by the model and is used in totality for testing purposes alone. This allows us to understand the generalization capability of COPS in real-time scenarios. COPS trained using this setup is represented by $COPS_{unseen\_test}$. No prior works use this setup; therefore, benchmarking is difficult. Nonetheless, this gives good insights.

		\textbf{\textit{Setup 3: Smishing} (S3):} We train and test the COPS architecture on the smishing data \cite{smishingdata}. As the dataset is tiny, we use the data generation framework. We train and test the model using synthetic and original data and benchmark the model against the current State-Of-The-Art (SOTA), which is trained in the same setup.

		\textbf{\textit{Setup 4: Smishing} (S4):}
		In this setup, we train COPS on the complete \cite{smishingdata} dataset and test on another unseen data \cite{kaggle}. Since there are no prior works, we train different model architectures like Random Forest, LSTM, VAE, etc., using the same setup and compare their performances with COPS.

		\begin{table}[h]
			\caption{Performance Comparison with Prior Works}
			\label{tab:metrics}
			\centering
			\resizebox{\linewidth}{!}{
				\begin{tabular}{ c|c|c|c|c|c }
					\toprule
					\rowcolor{lightyellow}
					\multicolumn{6}{c}{\textbf{URL Phishing}}\\ \cline{1-6}
					Setup &	Model & Accuracy & Precision & Recall & F1-Score\\
					\midrule
					\multirow{4}{*}{S1} &	PhishStorm \cite{marchal2014phishstorm} & 0.949 & 0.984 & -- & 0.947\\
					&	LSTM \cite{bahnsen2017classifying} & 0.987 & 0.986 & 0.989 & 0.987\\
					&	PhishHaven \cite{sameen2020phishhaven} & 0.98 & 0.98 & 0.979 & 0.97 \\
					&	\textbf{$COPS_{mix\_test}$} & \textbf{0.995} & \textbf{0.993} & \textbf{0.982} & \textbf{0.987}\\ \cline{1-6}
					S2	&	\textbf{$COPS_{unseen\_test}$} & 0.962 & 0.87 & 0.943 & 0.905 \\
					\midrule
					\rowcolor{lightyellow}
					\multicolumn{6}{c}{\textbf{Smishing+Spam}}\\ \cline{1-6}
					Setup	&	Model & Accuracy & Precision & Recall & F1-Score\\ \cline{1-6}
					
					\multirow{2}{*}{S3}	&	DSmishSMS \cite{mishra2021dsmishsms} & 0.979 & 0.84  & 0.94 & 0.887 \\
					&	\textbf{COPS} & \textbf{0.981} & \textbf{0.906}  & \textbf{0.962} & \textbf{0.934} \\
					
					\midrule
					\rowcolor{lightyellow}
					\multicolumn{6}{c}{\textbf{Smishing on Unseen Test Set}}\\ \cline{1-6}
					Setup	&	Model & Accuracy & Precision & Recall & F1-Score\\ \cline{1-6}
					
					\multirow{4}{*}{S4}	&	Random Forest  & 0.971 & 0.97  & 0.81 & 0.88 \\
					&	LSTM & 0.962 & 0.97  & 0.74 & 0.84 \\
					&	VAE & 0.974 & 0.96  & 0.84 & 0.90 \\
					&	\textbf{COPS} & \textbf{0.986} & \textbf{0.992}  & \textbf{0.904} & \textbf{0.945} \\
					\bottomrule
			\end{tabular}}
			\vskip -0.09in
		\end{table}
		
		\begin{figure}[t]
			\centering
			\includegraphics[trim={0cm 0cm 0cm 0cm},clip, width=0.7\textwidth, interpolate=false]{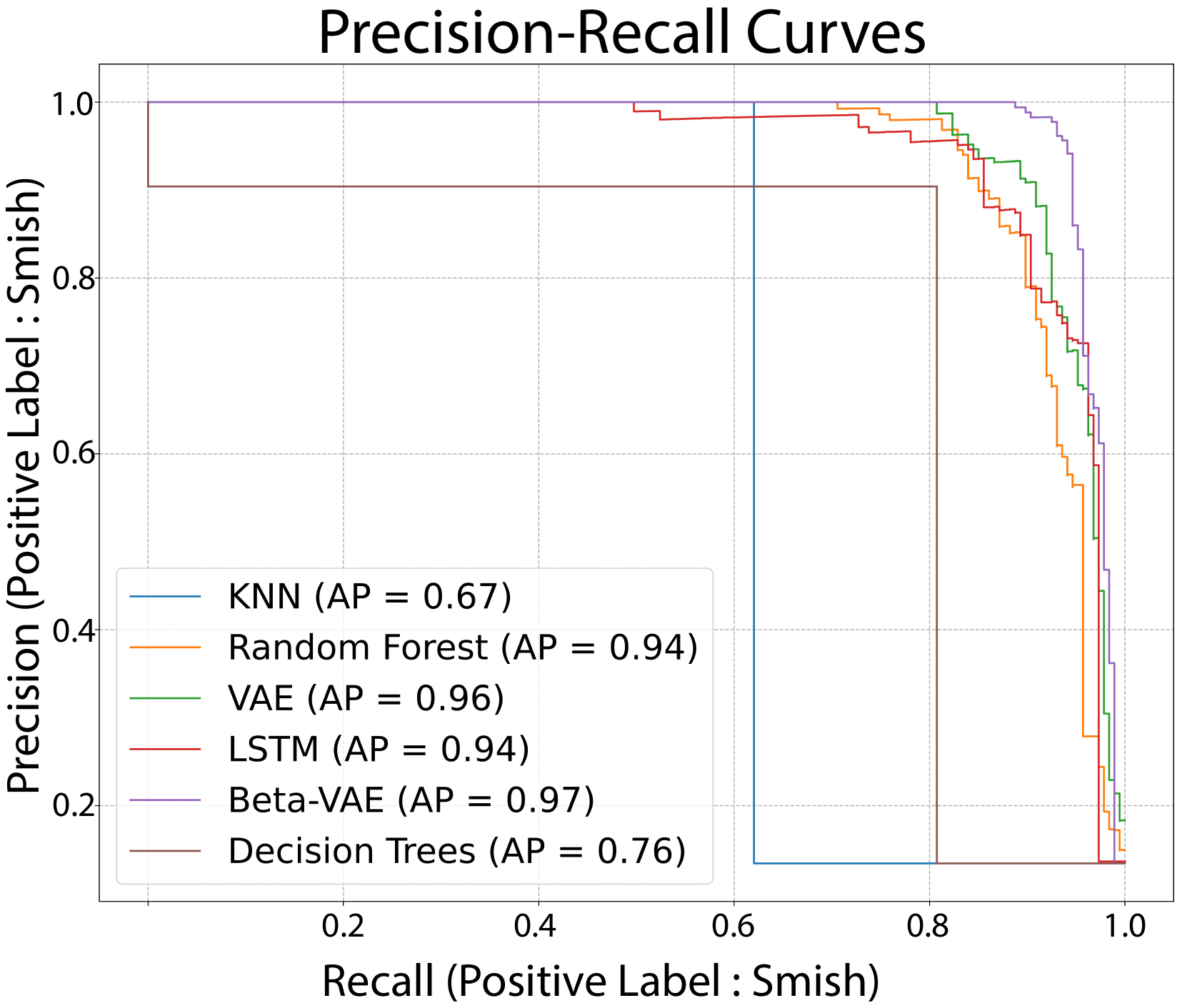}
			\caption{Precision-Recall curve showing comparative test results on Smishing with setup S4}
			\label{fig:curves}
			\vskip -0.15in
		\end{figure}

		\begin{figure}[h]
			\centering
			\begin{subfigure}{0.42\linewidth} 
				\includegraphics[width=\linewidth]{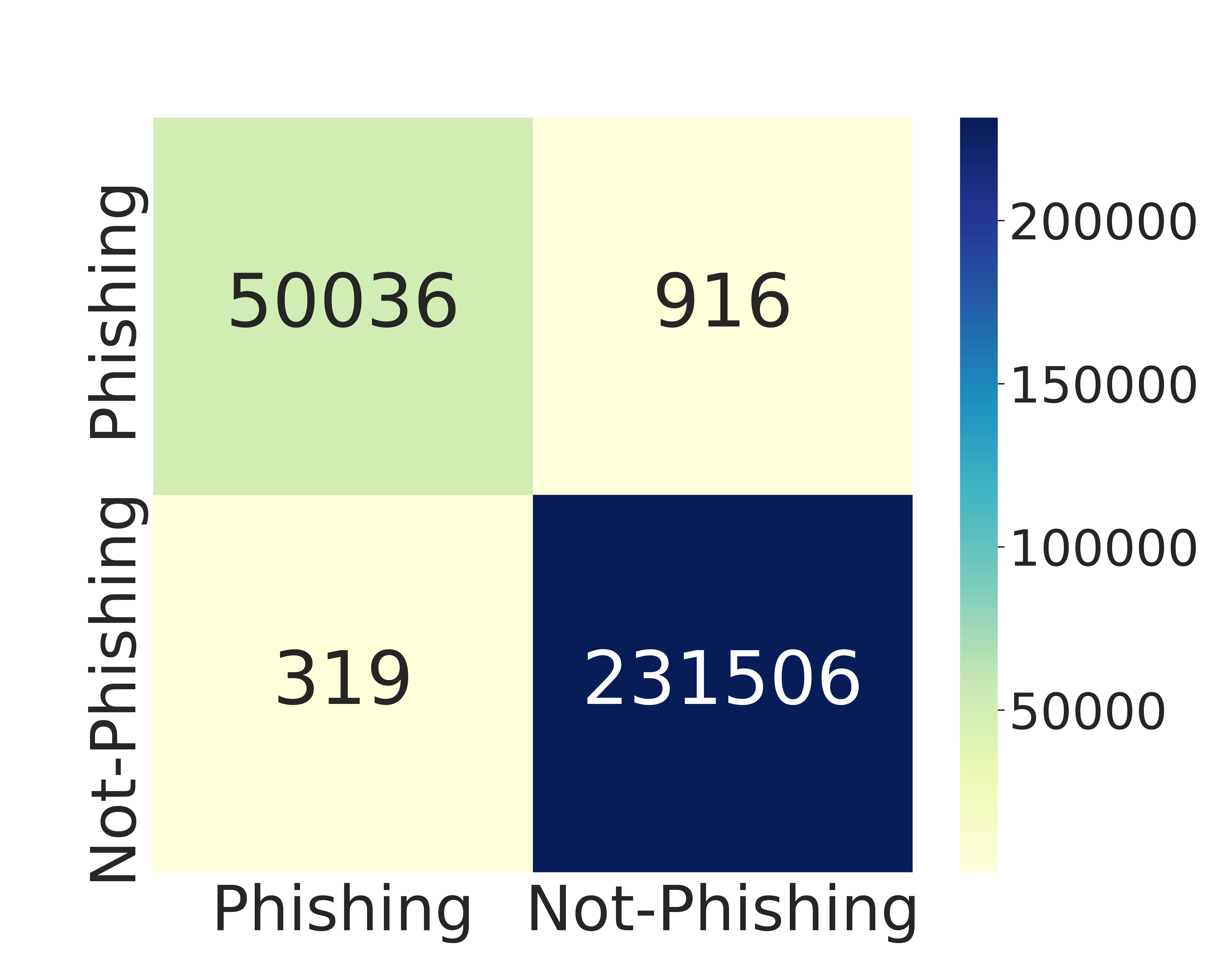}
				\caption{ \small $COPS_{mix\_test}$:Phishing}
				\label{fig:cm_url}
			\end{subfigure}
			\begin{subfigure}{0.42\linewidth} 
				\includegraphics[width=\linewidth]{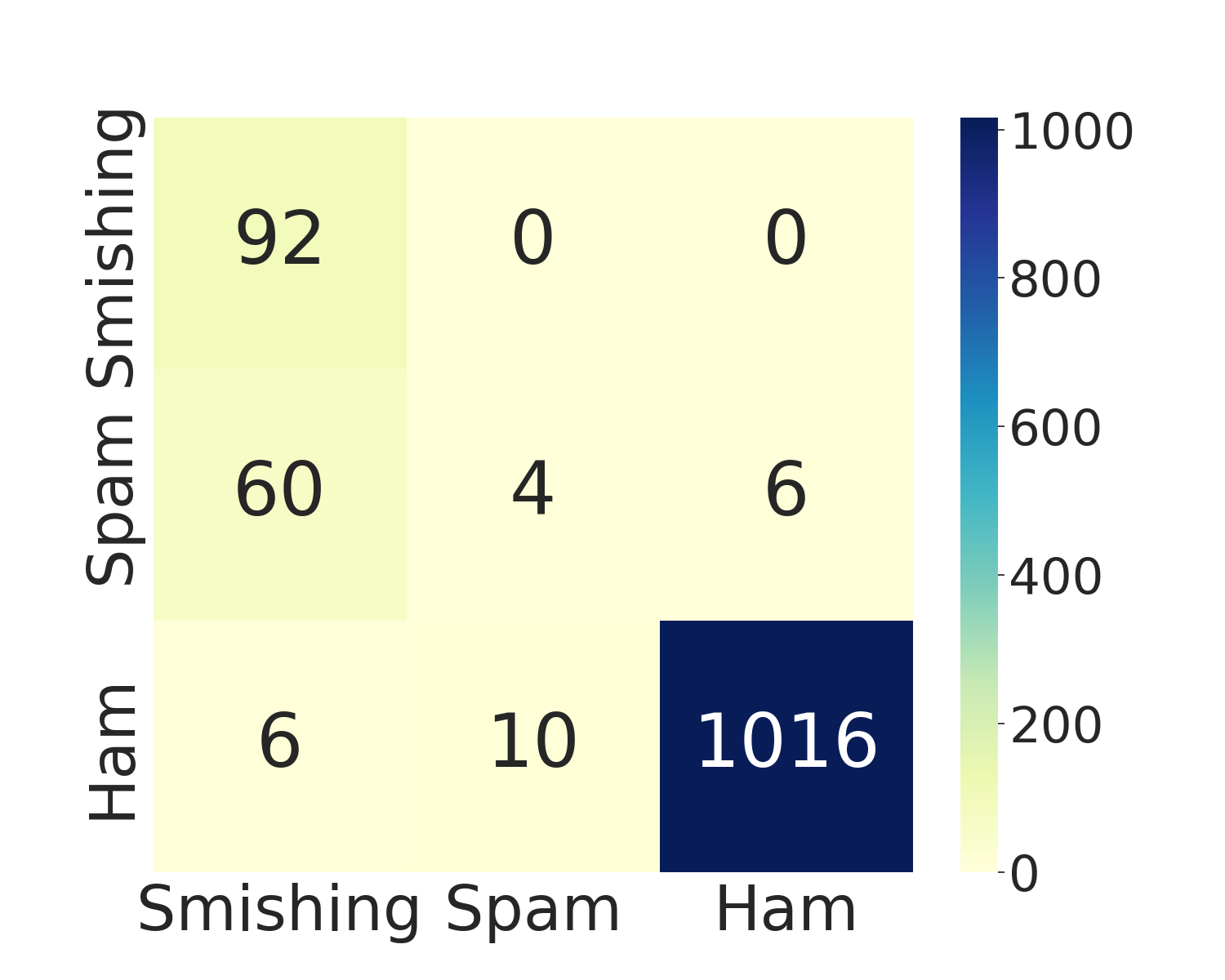} 
				\caption{\small $COPS$:Smishing}
				\label{fig:cm}
			\end{subfigure}
			\caption{Confusion matrices showing test results on a) URL Phishing (setup S1) and b) Smishing (setup S3). The x-axis represents predicted labels, and the y-axis the actual labels.}
			\label{fig:confusion_matrices}
		\end{figure}

		\subsection{Model Performance Evaluation}
		We evaluate the proposed architecture on Smishing and URL phishing tasks using different setups described previously on the open datasets mentioned in Section \ref{sec:DataCollection}.

		\textbf{URL Phishing:} The results of URL Phishing detection is shown in Table \ref{tab:metrics}. We test the performance of COPS using the S1 setup, which yields exemplary results with an accuracy of 99.5\% and an F1-Score of 98.7\%. The details of category-wise results are shown in the confusion matrix given in Figure \ref{fig:cm_url}.
		
		Test results of $COPS_{unseen\_test}$, which is trained according to setup S2, show a slight decrease in performance compared to $COPS_{mix\_test}$. This may be due to some completely unseen patterns in the test dataset. Nonetheless, The results are promising, with an F1-score of 90.5\%, assuring to deploy in real-world scenarios.
		
		We also benchmark on some prior works which perform URL phishing detection using models trained on the same PhishTank dataset \cite{phishstorm}. We can observe that our proposed model outperforms PhishStorm \cite{marchal2014phishstorm} and PhishHaven \cite{sameen2020phishhaven} in terms of all the metrics.

		\textbf{Smishing:} The results of setup S3 for smishing detection on the test set from the open dataset \cite{smishingdata} are shown in Table \ref{tab:metrics}. The details of category-wise results are shown in Figure \ref{fig:cm}. `Smishing' has a Recall of 100\%, which shows that our model predicts smishing very accurately. The performance observes a false positive rate of a mere 0.015 and a false negative rate of 0.037, combining both `Spam' and `Smishing' (reason for combining datasets explained in section \ref{sec:error}).  
		
		We benchmark the proposed model for Smishing detection on the DSmishSMS \cite{mishra2021dsmishsms} architecture owing to using the same dataset and set up for training and evaluation. Our model performs better in terms of accuracy and F1 score.
		
		To analyze the generalization ability, we use setup S4. COPS outperforms other architectures with an F1-score of 94.5\%. We further plot (Figure \ref{fig:curves}) the Precision-Recall curves to get deeper insights and compare COPS against other models. We can conclude that our model has good generalization capabilities.
		
		We also analyze the significance of synthetically generated data (further discussed in Section \ref{sec:ablation}). Data generations directly contribute to a 2.36\% rise in F1-Score.  
		
		\subsection{Error Analysis}
		\label{sec:error}
		
		\begin{figure}[h]
			\centering
			\includegraphics[width=0.75\linewidth]{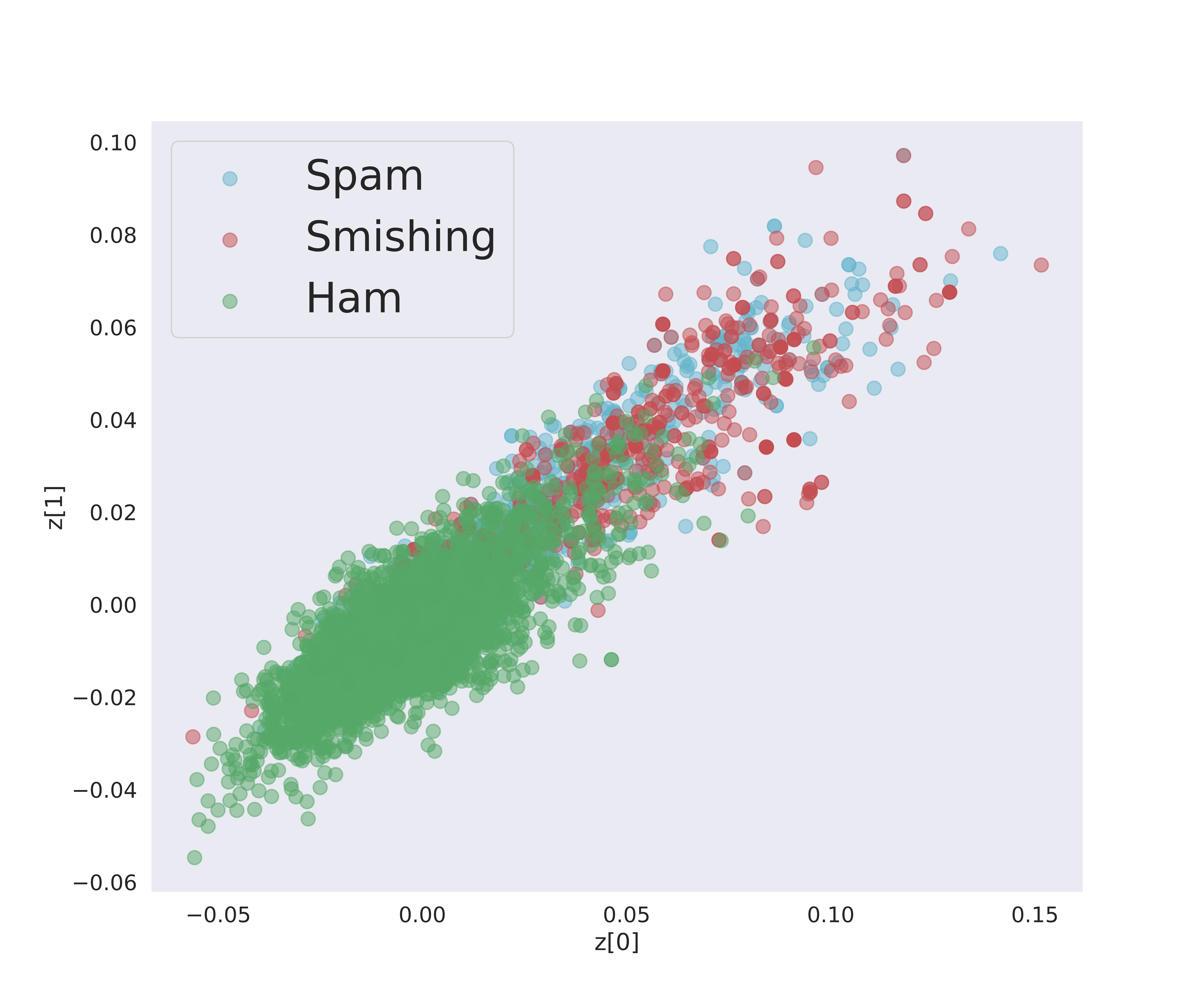}
			\caption{\textbf{Visualization of the Latent Distribution for Smishing}}
			\label{fig:graph_latent}
			\vskip -0.18in
		\end{figure}
		
		In the model performance results of setup S3, one notable drawback is that `Spam' messages are classified as `Smishing' for many test samples. To understand further, we plot the model's latent distribution of the dataset fit as shown in Figure \ref{fig:graph_latent} in which these two classes are highly overlapping. This can be seen from the dataset samples as well that the samples are very similar for `Spam' and `Smishing' and difficult to differentiate. Therefore, we design the real-time alerting system to alert the user if COPS predicts any input as either `Spam' or `Smishing'.
		
		A notable observation in the type of samples that are miss classified as `Ham' in setup S4 is that these samples contain a piece of fraudulent contact information with a text that looks legitimate. This issue can easily be tackled using FL based framework to keep track of fraudulent contact lists or integrate our solution with any existing open APIs that already keep track of these. We aim to work towards this in the future.

		\subsection{On-device Performance Evaluation}
		\label{sec:res_on-dev}
		We use the Samsung Galaxy S21 FE 5G smartphone model (Android 11.0, 6GB RAM, Snapdragon 888) to experiment and measure on-device performance. The proposed model requires minimal memory for processing ($\sim$ 12 MB) and storing the model itself (size of 3.46MB). We also evaluate model inference time achieving $\sim$ 0.08ms/char for URL phishing and smishing detection. These results further justify that our proposed architecture is optimal for deployment on resource-constrained devices.

		\subsection{Hyper-parameter Tuning}
		\label{sec:res_hyp}
\begin{figure}[h]
	\centering
	\includegraphics[width=0.73\linewidth]{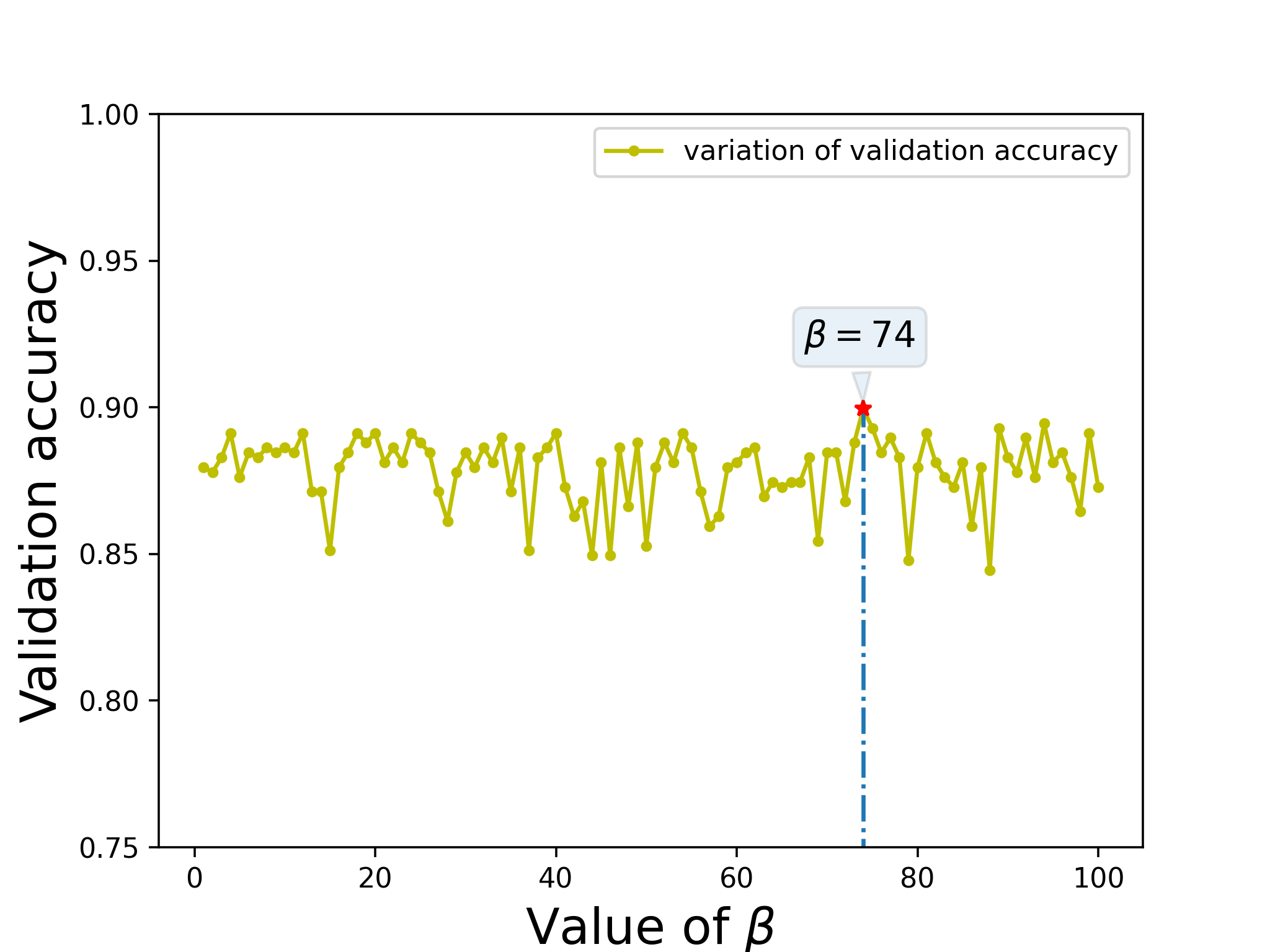}
	\caption{Variation of Accuracy with the hyper-parameter}
	\label{fig:beta}
	\vskip -0.15in
\end{figure}

		We also analyze the effect of hyperparameter tuning on the model performance with parameter $\beta$ for Smishing/URL Phishing detection. Figure \ref{fig:beta} shows the hyper-parameter tuning process of the value of $\beta$ for the most optimal results. We can see that we get the best results for the value $\beta = 74$.
		
		\subsection{Real-Time Performance Analysis}
		To test the performance of the deployed framework of COPS in real-time, we designed a user trial application. We distribute this application for testing, where the user is notified about detected smishing attacks and can reject the suggestion if it is wrong. We log the number of predictions across all test devices and the number of rejects across the users. We run this trial for 15 days across 50 users. We observe an accuracy of 94.92\% in detecting smishing attacks (256 predicted, 13 rejected). We plan to do an extended trial in the future for deeper insights and performance improvement.   
				\begin{table}[h]
			\caption{Ablation Study}
			\label{tab:ablation}
			
			\centering
			\resizebox{\linewidth}{!}{
				\begin{tabular}{ l|c|c|c }
					\toprule
					\rowcolor{lightyellow}
					\multirow{1}{*}{\textbf{Model}}                & 
					
					\multicolumn{3}{c}{\textbf{Smishing+Spam}}\\   \cline{2-4}
					\rowcolor{lightyellow}
					&  \textbf{Precision} & \textbf{Recall} & \textbf{F1 Score} 
					\\ \midrule
					
					Basic LSTM &  71.3   & 75.14   &  73.16       \\
						\hspace{0.3cm}+ Character Embedding    & 75.77  &  78.26 & 76.99        \\
						\hspace{0.6cm}+ BILSTM   & 78.82   & 83.75    & 81.21        \\
						\hspace{0.9cm}+ VAE   &  84.93    & 88.06  &   86.46        \\
						\hspace{1.2cm}+ $\beta$-VAE  &  89.34    & 92.81  &   91.04       \\
					\rowcolor{LightCyan}
						\hspace{1.5cm}+ Data Generation (COPS) & \textbf{90.69} & \textbf{96.29} & \textbf{93.40}\\ \bottomrule
			\end{tabular}}
		
		\end{table}
		
		\section{Ablation Study}
		\label{sec:ablation}
		We conduct an Ablation study to analyze the impact of each layer of the COPS detection module on performance. The details of this study are shown in Table \ref{tab:ablation}. We can see that character embedding help improve the performance with an increase of 3.83\% in the F1-score. This may be due to spelling errors and Personally Identifiable Information (PII), which is better modeled using character embeddings. We observe that using a VAE-based architecture results in a considerably significant addition in model performance with a 5.25\% rise F1-score. Disentangled VAE further improves the F1-score by 4.58\%. Training the model with augmenting synthetically generated data enhances the overall performance by a 2.36\%. This justifies the correctness and quality of the data generated using the proposed architecture.

		\section{Conclusion}
		Smishing is one of the significant problems smartphone users face today, and this issue is more dominant when the user is not tech-savvy. We propose COPS, a novel pipeline for real-time smishing detection, which can work across messaging/communication applications across different platforms. We also demonstrate that this complete pipeline can be implemented on-device with no interaction with any server elements, thus eliminating any concerns about compromising the privacy of the end users. Further, the COPS detection module comprising the Disentangled VAE-based architecture demonstrated that it can detect Smishing text with considerably good accuracy and validate the model performance on an open dataset. We also evaluate the model performance on URL phishing detection tasks.
		
		\bibliographystyle{IEEEtran}
		\bibliography{Short}

	\end{document}